\pdfoutput=1
\documentclass[sigconf]{acmart}

\usepackage{multirow}
\usepackage{algorithm,algorithmic}
\usepackage{geometry}
\AtBeginDocument{%
  \providecommand\BibTeX{{%
    \normalfont B\kern-0.5em{\scshape i\kern-0.25em b}\kern-0.8em\TeX}}}

\setcopyright{acmlicensed}
\copyrightyear{2018}
\acmYear{2018}
\acmDOI{XXXXXXX.XXXXXXX}

\acmConference[Conference acronym 'XX]{Make sure to enter the correct
  conference title from your rights confirmation emai}{June 03--05,
  2018}{Woodstock, NY}
\acmISBN{978-1-4503-XXXX-X/18/06}

\begin{document}

\title{Hypergraph Enhanced Knowledge Tree Prompt Learning for Next-Basket Recommendation}

\author{Zi-Feng Mai}
\email{maizf3@mail2.sysu.edu.cn}
\affiliation{%
  \institution{Sun Yat-Sen University}
  \city{Guangzhou}
  \state{Guangdong}
  \country{China}
}
\author{Chang-Dong Wang}
\authornote{Corresponding author}
\email{changdongwang@hotmail.com}
\affiliation{%
  \institution{Sun Yat-Sen University}
  \city{Guangzhou}
  \state{Guangdong}
  \country{China}
}

\author{Zhongjie Zeng}
\email{zengzhjgdut@gmail.com}
\affiliation{%
  \institution{Wens Foodstuff Group Co., Ltd.}
  \city{YunFu}
  \state{Guangdong}
  \country{China}
}

\author{Ya Li}
\email{liya2829@gpnu.edu.cn}
\affiliation{%
  \institution{Guangdong Polytechnic Normal University}
  \city{Guangzhou}
  \state{Guangdong}
  \country{China}
}

\author{Jiaquan Chen}
\email{18998812918@189.cn}
\affiliation{%
  \institution{Wens Foodstuff Group Co., Ltd.}
  \city{YunFu}
  \state{Guangdong}
  \country{China}
}

\author{Philip S. Yu}
\email{psyu@cs.uic.edu}
\affiliation{%
  \institution{University of Illinois}
  \city{Chicago}
  \state{Illinois}
  \country{U.S.A.}
}

\renewcommand{\shortauthors}{Mai, et al.}

\begin{abstract}

Next-basket recommendation (NBR) aims to infer the items in the next basket given the corresponding basket sequence. Existing NBR methods are mainly based on either message passing in a plain graph or transition modelling in a basket sequence. However, these methods only consider point-to-point binary item relations while item dependencies in real world scenarios are often in higher order. Additionally, the importance of the same item to different users varies due to variation of user preferences, and the relations between items usually involve various aspects. As pretrained language models (PLMs) excel in multiple tasks in natural language processing (NLP) and computer vision (CV), many researchers have made great efforts in utilizing PLMs to boost recommendation. However, existing PLM-based recommendation methods degrade when encountering Out-Of-Vocabulary (OOV) items. OOV items are those whose IDs are out of PLM's vocabulary and thus unintelligible to PLM. 
To settle the above challenges, we propose a novel method HEKP4NBR, which transforms the knowledge graph (KG) into prompts, namely Knowledge Tree Prompt (KTP), to help PLM encode the OOV item IDs in the user's basket sequence. A hypergraph convolutional module is designed to build a hypergraph based on item similarities measured by an MoE model from multiple aspects and then employ convolution on the hypergraph to model correlations among multiple items. Extensive experiments are conducted on HEKP4NBR on two datasets based on real company data and validate its effectiveness against multiple state-of-the-art methods. 
  
\end{abstract}




\received{20 February 2007}
\received[revised]{12 March 2009}
\received[accepted]{5 June 2009}

\maketitle

\section{Introduction}


Recently, with the explosive growth in e-commerce and online platforms, recommendation algorithms have become a potential research field that attracts much attention in both academia and industry. One of the most popular topics is called Next-Basket Recommendation (NBR), which groups multiple items interacted at the same time into a basket and aims to infer the items in the basket that user will interact with given his/her basket sequence. Compared with Sequential Recommendation (SR) \cite{gru4rec,bert4rec}, NBR has a wider application in real world scenarios since SR assumes that user interactions within a period of time must strictly follow a chronological order. This assumption is difficult to maintain since users often interact with multiple items simultaneously to satisfy a certain need. 

A common issue of NBR task is data sparsity, meaning that each user only interacts with a small portion of items, resulting in insufficient data for learning representations of users and items \cite{hapcl}. To tackle this issue, some existing NBR methods focus on mining sequential transition patterns in the historical interaction sequences \cite{WuPersonalized2022,M2,Kou2023modeling,deng2023multi,wang2020Intention}. However, these methods rely solely on item IDs, which leads to underfitting in the representation and is further expressed as a degradation of recommendation performance especially for the long-tail items. Some other NBR methods attempt to mine collaborative signals by considering user interactions as graphs \cite{bpcl-tingting,hapcl,digbot,brl}. These methods model the correlations between two items by passing messages along the edges and aggregating messages from the neighbours. However, due to the data sparsity issue, those graphs often contain only shallow collaborative signals, making it difficult to mine deep semantic correlations. Furthermore, items that the user haven't interacted with do not necessarily mean that these items do not match the user's preferences.

Despite of the remarkable improvements that these methods have achieved, they still face the following challenges. 
\begin{itemize}
    \item 
    Existing methods are mainly based on either message passing in a plain graph or transition modelling in an interaction sequence. These methods only model point-to-point binary item relation, whereas item dependencies are beyond binary relation and are often triadic or even higher order. For example, many widely used therapies for diseases such as HIV and pulmonary tuberculosis often involve a combination of multiple medicines. 
    \item 
    Due to variations in user preferences, the importance of items varies. For example, for users with sufficient budgets, the quality of items is often the primary consideration. However, for those on a tight budget, price is also an important factor. What's more, the relationships between items involve various aspects, such as complementarity, substitution and cooperation. 
    \item 
    Some methods attempt to incorporate side information such as KG to alleviate the data sparsity issue \cite{gru4recf,kgat,ckan}. However, obtaining high quality item representations in large scale KG is a challenging task due to the intrinsic noise in the KG. For example, \textit{"Insulin is a medication for diabetes"} has nothing to do with \textit{"Penicillin is a medication for bacterial infections"}. However, they will infect each other through the message passing process in KG. Furthermore, the pre-trained item representations may not be suitable for the downstream task. 
\end{itemize}

Recently, PLMs have shown remarkable performance in NLP \cite{gpt3,t5,llama} and CV \cite{clip,dall-e,glip}, demonstrating strong capabilities in comprehension and representation. As PLMs are applied to various tasks, a new learning paradigm called prompt learning has emerged. Prompt learning follows the \textit{"pretrain-prompt-tuning"} paradigm, which transforms downstream tasks into pre-train tasks of PLMs using prompt templates to bridge the gap between the downstream objectives and pre-train objectives, seamlessly transferring prior knowledge in PLMs to downstream tasks. There are many research on utilizing PLMs in recommendation \cite{sun2023generative,Li2023Personalized,Wang2022Towards,zhai2023knowledge,zhang2023prompt}. 

\begin{figure}
	\centering
		\includegraphics[width=0.8\linewidth]{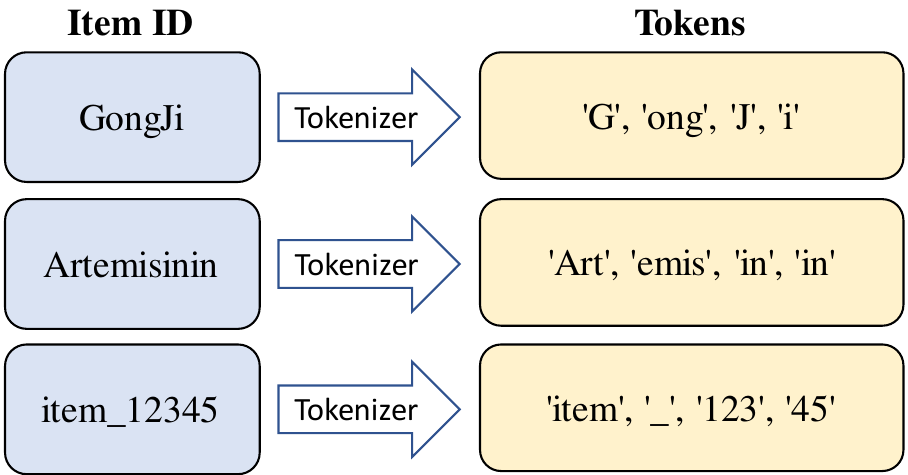}
		\caption{Example of OOV item IDs. These IDs become irrelevant tokens after tokenization by tokenizers. We use T5-small \cite{t5} tokenizer in this example.}
	\label{fig:oov}
\end{figure}

However, PLM-based recommendation is not as straightforward as described above. One of the biggest challenges is how to represent items in prompt. Since the PLM corpus cannot be infinite, it often struggles to understand words that are not present in its corpus. We define these items as Out-Of-Vocabulary (OOV) items, meaning that their item IDs are out of PLM's vocabulary. There are mainly two ways in existing methods to represent an item. One way is to directly use its original name but many OOV item names are either terminologies (such as a medicine called \textit{Artemisinin}) or non-English words (such as a Chinese poultry called \textit{GongJi}). The other way is to map each item to a unique number and treat that number as its item ID (such as \textit{item\_12345}). As \autoref{fig:oov} shows, neither of these two ways is intelligible to PLMs. For instance, \textit{Artemisinin} has nothing to do with token \textit{Art} and token \textit{123} is not semantically correlated with the following token \textit{45}. 


In order to tackle the aforementioned challenges, we propose \textbf{H}ypergraph \textbf{E}nhanced \textbf{K}nowledge Tree \textbf{P}rompt Learning for \textbf{N}ext-\textbf{B}asket \textbf{R}ecommendation (HEKP4NBR) in this paper. First, inspired by P5 \cite{p5}, we propose to construct a masked user prompt (MUP) according to the user's basket sequence, where each $[MASK]$ token in MUP corresponds to an item in the next basket. In this way, we can transform the NBR task of predicting items in the next basket into a pre-train task called mask prediction which predicts tokens corresponding to the $[MASK]$ tokens. Then, in order to help PLMs encode the OOV items in MUP, we propose to construct knowledge tree prompts (KTP) by building a knowledge tree based on user's historical interaction and KG inspired by \cite{zhai2023knowledge}. In this way, we are able to provide rich information of those OOV items while avoiding brutal encoding of the whole KG to mitigate the noise issue. What's more, inspired by several works on hypergraphs \cite{yang2022multi,digbot,brl}, we design a multi-item relation encoder, which employs an MoE model to measure item similarity from multiple aspects and builds a hypergraph based on item similarity. Then we use a hypergraph convolutional module to model the correlations among multiple items. Lastly, we use a simple yet effective frequency base gating module to dynamically matching basket sequence to each item by the frequency of the user interaction. Extensive experiments show that our method outperforms state-of-the-art methods on multiple metrics in two real world datasets. The contributions of this paper can be summarized as follows:
\begin{itemize}
    \item We proposed HEKP4NBR which mines the transition pattern in the basket sequence with the help of PLMs and KG while utilizing hypergraph to model the correlations among multiple items.
    \item We construct MUP to transform NBR task into mask prediction task following the prompt learning paradigm. In order to encode the OOV items in MUP, we construct another prompt KTP to leverage knowledge in KG.
    \item We design a multi-item relation encoder model the correlations among multiple items with an MoE model and a hypergraph convolutional module.
    \item We conduct extensive experiments on two datasets, Poultry and Pharmacy. They are based on two real world companies providing B2B and B2C service each. The results demonstrate the effectiveness of our method. 
    
\end{itemize}

\section{Related Work}

\subsection{Next-Baket Recommendation}
Next-basket recommendation aims at inferring multiple items in the next basket. Early works \cite{Rendle2010factorizing,wan2015next} utilize Markov Chain (MC) which strictly modelling transaction between every two nearby baskets, limiting the exploration of user interest. With the development of deep learning, RNN-based methods such as DREAM \cite{yu2016a}, CLEA \cite{clea} and Beacon \cite{le2019correlation} apply RNN to capture the long-term dependency in basket sequence. Similarly, some attention-based methods \cite{wang2020Intention,wang2020Intention2Basket} adaptively mine the different contributions of each item to user preferences using attention mechanism. Recently, $M^2$ \cite{M2} takes the users general preferences, items global popularities and sequential transition patterns among items into consideration. SecGT \cite{Kou2023modeling} models user preference by collaboratively modeling the joint user-seller-product interactions and sequentially exploring the user-agnostic basket transitions. MMNR \cite{deng2023multi} normalizes the number of interactions from both user and item views to obtain differentiated representation and aggregates item representations within baskets from multi aspects. While these methods have produced competitive results, relying solely on item IDs leads to the model treating each item independently. Consequently, the model can only capture shallow ID transition patterns and lack the ability to discern deeper semantic-level correlations.

\subsection{Graph-Based Recommendation}


Graph is a data structure that represents the associations between entities by connectivity and there's consistency between graph structure and user interaction. Early works \cite{deepwalk,node2vec} employ random walk to extract some node sequences in graph and encode each node in an embedding. Some fundamental works such as GCN \cite{gcn}, GAT \cite{gat}, GraphSage\cite{graphsage} and LightGCN \cite{lightgcn} design various GNNs for learning representations of nodes in graph, after which KGCN \cite{kgcn} and KGAT \cite{kgat} extend these works to Knowledge Graph (KG). Recently, GCSAN \cite{gcsan} uses self-attention layers to assign different weights for each item by constructing a directed graph from item sequences. SINE \cite{sine} mines multiple aspects of interest from item sequence to effectively model several diverse conceptual prototypes. These methods only model point-to-point binary item relation while in real world scenarios, especially in B2B scenarios, item dependencies are often higher order. To tackle this issue, DigBot \cite{digbot} constructs a hypergraph that connects each item in a basket by a hyperedge. BRL \cite{brl} employs hypergraph convolution on hypergraph to explore correlations among different items. However, these methods fail to mine deep semantic correlations since the hypergraphs are built only based on the inclusion relationship between baskets and items.

\subsection{PLM for Recommendation}


PLMs have achieved remarkable success in various tasks, and there is a large number of PLM-based recommendation methods have been proposed recently. PEPLER \cite{pepler} uses GPT-2 as an interpreter to give personalized explanations for recommendation results. NRMS \cite{NRMS} serves PLMs as item encoders for news recommendation and Prompt4NR \cite{zhang2023prompt} designs several discrete and continuous prompt templates following prompt learning paradigm. P5 \cite{p5} is a unified framework that integrates several recommendation tasks by transforming them into NLP tasks. Following P5, KP4SR \cite{zhai2023knowledge} further incorporates KG into prompt and use masks to block attention between irrelevant triplets. GenRec \cite{sun2023generative} generate each item in the next basket one by one in an auto-regressive manner to simulate the process of users satisfying their needs step by step. However, none of these works take into consideration how items are represented in prompts, resulting in performance degradation when encountering OOV items.

\section{Methodology}

\subsection{Problem Formulation and Preliminary}
We first give a formulation on Next-Basket Recommendation. Let $\mathcal{U}=\{u_1,\dots,u_{|\mathcal{U}|}\}$ denotes the user set and $\mathcal{I}=\{i_1,\dots,i_{|\mathcal{I}|}\}$ denotes the item set. A basket $b=\{i_1^b,\dots, i_{|b|}^b\}$ is a set of $|b|$ items which are interacted by a user $u\in\mathcal{U}$ at the same time.
Let $\mathcal{B}=\{b_1,b_2,\dots,b_{|\mathcal{B}|}\}$ denotes the set of all basket. The history interaction sequence of a user $u\in\mathcal{U}$ can be denoted as a basket sequence arranged in chronological order, i.e.,  $\mathcal{S}^u=\left[b_1^u,b_2^u,\dots,b_{|\mathcal{S}^u|}^u\right]$ where $b_i^u\in\mathcal{B}$. The NBR task is to predict the next basket $b_{|\mathcal{S}^u|+1}^u$ that user $u$ may interact given his basket sequence $\mathcal{S}^u$. 

Then, we provide a definition on knowledge graph (KG). A KG is a heterogeneous graph consists of an entity set $\mathcal{E}$ and a relation set $\mathcal{R}$, which can be defined as $\mathcal{KG}=\{(h, r, t)\mid h,t\in\mathcal{E}, r\in\mathcal{R}\}$. A triplet $(h,r,t)$ indicates that relation $r$ connects the head entity $h$ and the tail entity $t$. 

Next, we define the basket-item bipartite graph as a bipartite graph between basket set $\mathcal{B}$ and item set $\mathcal{I}$, denoted by $\mathcal{G}_{B-I}=(\mathcal{B}, \mathcal{I}, E_{B-I})$, where each edge $(b,i)$ in edges set $E_{B-I}$ indicates that item $i\in\mathcal{I}$ is contained in basket $b\in\mathcal{B}$.

Lastly, we briefly introduce hypergraph. An hypergraph $\mathcal{HG}=(V,E)$ consists of a vertex set $V$ and a hyperedge set $E$. Each hyperedge $e\in E$ is a subset of $V$, indicating that multiple vertices are connected by $e$. The degree of a vertex $v$, denoted by $d_v$, is the number of hyperedges that $v$ is incident to. Similarly, the degree of a hyperedge $e$, denoted by $d_e$, is the number of vertices that are incident to $e$. 

\subsection{Overview}

\begin{figure*}
	\centering
		\includegraphics[width=0.9\linewidth]{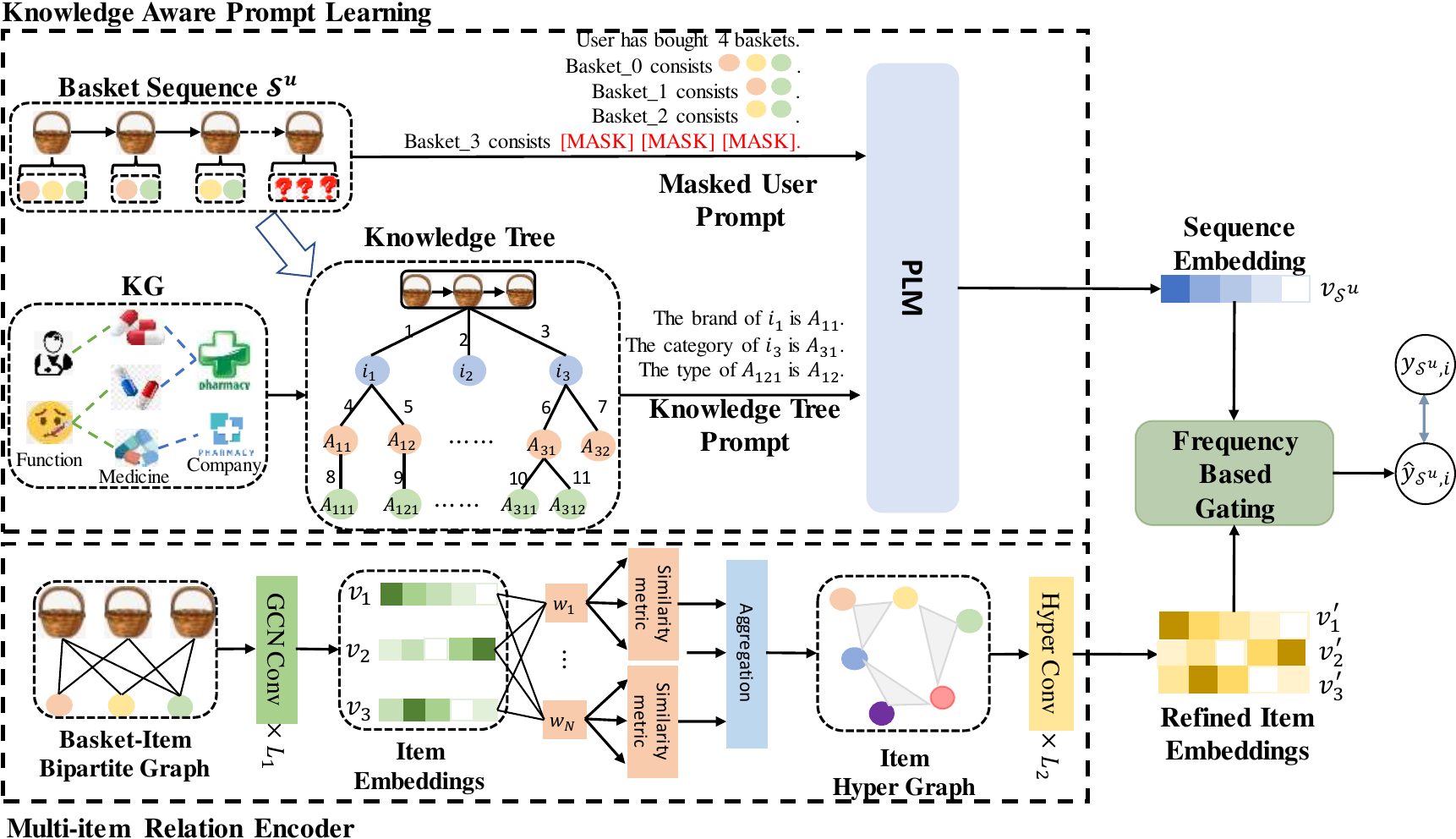}
		\caption{The framework of proposed HEKP4NBR model. We first construct MUP and KTP base on user's basket sequence $\mathcal{S}^u$ and KG to obtain the sequence embedding $\mathbf{v}_{\mathcal{S}^u}$ by PLM. Then, the refined embeddings of each item $i\in\mathcal{I}$, denoted by $\mathbf{v}_i'$, are obtained by the multi-item relation encoder, more specifically, MoE model and hypergraph convolutional module. Lastly, we dynamically match $\mathbf{v}_{\mathcal{S}^u}$ to each $\mathbf{v}_i'$ by frequency based gating mechanism to obtain the probability that item $i$ consists in the next basket, denoted by $\hat{y}_{\mathcal{S}^u,i}$. Besides from the traditional cross-entropy loss $\mathcal{L}_{Rec}$ and auto-regressive loss $\mathcal{L}_{PLM}$, we also design two pair-wise ranking loss of two different levels, namely basket-item level loss $\mathcal{L}_{B-I}$ and item-item level loss $\mathcal{L}_{I-I}$.}
	\label{fig:model}
\end{figure*}

We propose a model called HEKP4NBR, which transforms KG into prompt to alleviate the ambiguity when PLM modelling OOV items and models the correlation among multiple items by hypergraph convolutional operation. It mainly consists of three modules: knowledge aware prompt learning module (Section~\ref{module:prompt}), multi-item relation encoder module (Section~\ref{module:hypergraph}) and frequency based gating module (Section~\ref{module:fbg}). The overall framework and pipeline of HEKP4NBR is shown in \autoref{fig:model}.

\subsection{Knowledge Aware Prompt Learning}
\label{module:prompt}
The \textit{pretrain-prompt-tuning} paradigm has achieved great success in NLP and CV, and many efforts have been made to apply prompt learning in recommender systems \cite{p5,zhai2023knowledge}. In this section, we construct two kinds of prompt, namely masked user prompt (MUP) and knowledge tree prompt (KTP), which describe the interacted basket sequence of a user and knowledge of items in the basket sequence respectively. Combining MUP and KTP, we can not only utilize generic knowledge in PLM, but also alleviate the detriment of OOV items on PLM.

\subsubsection{Masked user prompt}
Inspired by \cite{p5}, we construct a masked user prompt to transform NBR task into a pre-train task, i.e., mask prediction. In this way, our model can bridge the gap between pre-train task and downstream task.

Specifically, we manually design some templates $\mathcal{T}_{MUP}$ to transform $\mathcal{S}^u$ into MUP, that is,
\begin{equation}
    \text{MUP}=\mathcal{T}_{MUP}(\mathcal{S}^u)
    \label{formula:mup}
\end{equation}
For instance, suppose the basket sequence of user $u$ is $\mathcal{S}_u=\left[\{ A, B \} , \{ B , C \} , \{ C , D \}\right]$. Its corresponding MUP is \textit{"User $u$ has purchase 3 baskets. Basket\_0 consists of $A$, $B$. Basket\_1 consists of $B$, $C$. Basket\_2 will consists of $[MASK]_1$, $[MASK]_2$"}, where $[MASK]_1$ and $[MASK]_2$ correspond to items in the next basket, i.e., $C$ and $D$.

As mentioned above, when faced with OOV items, these words may not be as simple as $A$, $B$ or $C$, but words that are beyond PLM's knowledge, or even non-English words. In this scenario, we argue that these unintelligible words would do harm to PLM while introducing side information such as KG will help. Therefore, we also construct the following KTP to incorporate KG into prompts.

\subsubsection{Knowledge tree prompt}
\label{module:ktp}
Many research \cite{wang2019multi,wang2021learning} has shown that introducing KG as side information in recommendation may significantly boost performance. However, as mentioned before, it's still a open problem about how to corporate KG into prompts since brutally flattening structured KG into linear words would lead to inevitable structural information loss. In order to settle this challenge, we proposed to readout KTP from KG by constructing a knowledge tree based on basket sequence $\mathcal{S}^u$.

In order to build the knowledge tree, we first construct the augmented KG, denoted by $\mathcal{KG}'$, following \autoref{formula:aug_kg},
\begin{equation}
    \mathcal{KG}'=\mathcal{KG}\cup\bigcup_{i\in\mathcal{S}^u}\{\left(s, \textit{consist\_of},  i\right)\}
    \label{formula:aug_kg}
\end{equation}
where $s$ is a new entity representing the basket sequence $\mathcal{S}^u$ and \textit{consist\_of} is a new relation. Then we perform beam search in $\mathcal{KG}'$ starting from the new entity $s$ to all entities within $n$ hops from $s$, after which the beam search tree is called as knowledge tree. Next, we traverse knowledge tree in breadth-first manner and sequentially record every triplet during the traversal to create a triplet sequence $\mathcal{S}^{tri}$,
\begin{equation}
\mathcal{S}^{tri}=\left[\left(s,\textit{consist\_of}, i_1\right),\left(s,\textit{consist\_of}, i_2\right) ,\dots,\left(h_L,r_L, t_L\right)\right]
\label{formula:knowledgetree}
\end{equation}
where $L$ is the number of all triplets, noted that each triplet $(h,r,t)\in\mathcal{S}^{tri}$ indicates an edge of relation $r$ points from head entity $h$ to tail entity $t$ in $\mathcal{KG}'$. The KTP can be defined as \autoref{formula:ktp},

\begin{equation}
    \label{formula:ktp}
    \text{KTP} = \text{concat}(\mathcal{T}_{KTP}(h_1,r_1,t_1),\dots,\mathcal{T}_{KTP}(h_L,r_L,t_L))
\end{equation}
where $\mathcal{T}_{KTP}$ is the KTP template that converts a triplet into a string. In this paper, we define $\mathcal{T}_{KTP}(h,r,t)=\textit{"The r of h is t"}$. 

\subsubsection{PLM encoding and fine-tuning}
After we construct MUP and KTP base on user's basket sequence $\mathcal{S}^u$ and KG, we input them into PLM and obtain the output embeddings of $[MASK]$ tokens. Noted that there might be multiple $[MASK]$ tokens, we just simply employ a mean pooling over all $[MASK]$ embeddings to get the sequence embedding $\mathbf{v}_{\mathcal{S}^u}\in\mathbb{R}^{d_1}$, where $d_1$ denotes hidden size of PLM.

\begin{equation}
    \label{formula:seq_emb}
    \mathbf{v}_{\mathcal{S}^u}=\text{MeanPool}(PLM_{[MASK]}(MUP,KTP))
\end{equation}

Following \cite{p5}, we use the auto-regressive Negative Log Likelihood (NLL) loss $\mathcal{L}_{PLM}$ to supervise PLM fine-tuning following \autoref{fomula:plm_loss},

\begin{equation}
    \mathcal{L}_{PLM}=-\sum_{j=1}^{|\mathbf{y}|}\log{P_\Theta(\mathbf{y}_j\mid \mathbf{y}_{<j},\mathbf{X}_\text{MUP},\mathbf{X}_\text{KTP})}
    \label{fomula:plm_loss}
\end{equation}
where $\mathbf{y}$ represents the predicted output and $\Theta$ represents the parameters in PLM.

\subsection{Multi-item Relation Encoder}
\label{module:hypergraph}



\subsubsection{Item Semantic Encoder with MoE model}

We argue that there is a greater similarity between items within the same basket. Thus, we conduct $L_1$ layers of GCN convolution to the basket-item bipartite graph $\mathcal{G}_{B-I}$ to obtain an initial embedding for each item $i\in\mathcal{I}$, denoted by $\mathbf{v}_i\in\mathbb{R}^{d_2}$, and each basket $b\in\mathcal{B}$, denoted by $\mathbf{v}_b\in\mathbb{R}^{d_2}$, where $d_2$ represents the size of embeddings. 

In order to encode the correlations and dependencies among items, we introduce an item semantic encoder to obtain the item similarity matrix $\mathbf{\Pi}\in\mathbb{R}^{|\mathcal{I}|\times|\mathcal{I}|}$, whose $(i,j)$-th entry $\pi_{i,j}$ represents the similarity between item $i$ and item $j$. We employ a Mixture-of-Experts, or MoE, model to learn $\pi_{i,j}$ from multiple aspects following \autoref{formula:moe1} and \autoref{formula:moe2},
\begin{align}
    \label{formula:moe1}
    \pi_{i,j}^n&=\tau(w_n\mathbf{v}_i,w_n\mathbf{v}_j)\\
    \label{formula:moe2}
    \pi_{i,j}&=Agg(\pi_{i,j}^1,\pi_{i,j}^2,\dots,\pi_{i,j}^N)
\end{align}
where $N$ represents the number of experts and $w_n\in\mathbb{R}^{d_2\times d_3}$ is learnable parameter. $\tau:\mathbb{R}^{d_3}\times\mathbb{R}^{d_3}\mapsto\mathbb{R}$ is a similarity metric function. $Agg$ is an aggregating function that aggregate from all experts to obtain the final similarity.

We design two pair-wise ranking losses of two different levels. The first ranking loss, denoted $\mathcal{L}_{B-I}$, is based on the hypothesis that the embedding of a basket $b$ should be more similar with the embedding of an item $i^+$ that includes in basket $b$ than it of an item $i^-$ that excludes. Given a basket $b$, we randomly sample a positive item $i^+\in b$ and another negative item $i^-\notin b$ then compute $\mathcal{L}_{B-I}$ as \autoref{formula:gcn_loss},
\begin{equation}
    \label{formula:gcn_loss}
    \mathcal{L}_{B-I}=-\sum_{b\in\mathcal{B}}\log\left(\sigma\left(\mathbf{v}_b\cdot \mathbf{v}_{i^+}-\mathbf{v}_b\cdot \mathbf{v}_{i^-}\right)\right)
\end{equation}
where $\cdot$ represents the inner product of vectors and $\sigma$ is the sigmoid function. 

The second ranking loss, denoted $\mathcal{L}_{I-I}$, is based on the hypothesis that the similarity of an item pair $(i,i^+)$ that appears in the same basket should be higher than it of an item pair $(i,i^-)$ that doesn't. Given a basket $b$ and for each item $i\in b$, we randomly sample an positive item $i^+\in b/\{i\}$ and another negative item $i^-\notin b$ and compute $\mathcal{L}_{I-I}$ as \autoref{formula:hgcn_loss},
\begin{equation}
    \label{formula:hgcn_loss}
    \mathcal{L}_{I-I}=-\sum_{b\in\mathcal{B}}\frac{1}{|b|}\sum_{i\in b}\log\left(\sigma\left(\pi_{i,i^+}-\pi_{i,i^-}\right)\right)
\end{equation}
where $|b|$ represents the size of basket $b$. 

\subsubsection{Item hypergraph Convolution}
With the item similarity matrix $\mathbf{\Pi}$ encoded by the MoE model, we then construct a item hypergraph $\mathcal{HG}$ by connecting each item $i$ with its top-$k$ most similar items using a hyperedge. Specifically, given the item similarity matrix $\Pi$, we define the adjacent matrix of item hypergraph as $\mathcal{M}\in\mathbb{R}^{|\mathcal{I}|\times|\mathcal{I}|}$, whose $(i,j)$-th entry $m_{i,j}$ is calculated by \autoref{formula:adj_matrix},
\begin{equation}
    \label{formula:adj_matrix}
    m_{i,j}=
    \begin{cases}
        \pi_{i,j}& j\in \mathcal{A}_i\\
        0&otherwise
    \end{cases}
\end{equation}
where $\mathcal{A}_i$ represents the set of the top-$k$ similar items of item $i$ according to $\Pi$. 

Inspired by \cite{yang2022multi}, we design a hypergraph convolution module by applying message passing paradigm on hyperedges to capture the correlations among multiple items. Specifically, we design our hypergraph convolutional layer as \autoref{formula:hyper_conv},
\begin{equation}
    \label{formula:hyper_conv}
    \mathbf{H}^{(l+1)}=\mathrm{FFN}\left(\mathbf{D}_v^{-1}\cdot\mathcal{M}\cdot\mathbf{D}_e^{-1}\cdot\mathcal{M}^\top\cdot\mathbf{H}^{(l)}\right)
\end{equation}
where $\mathbf{D}_v$ and $\mathbf{D}_e$ denotes the diagonal matrices composed of the degrees of each nodes and hyperedges in the hypergraph $\mathcal{HG}$ respectively. $\mathrm{FFN}$ represents the learable feed forward layer. Each embedding $\mathbf{h}_i^{(0)}$ is initalized by the former GCN convolutional layers, i.e., $\mathbf{h}_i^{(0)}=\mathbf{v}_i\in\mathbb{R}^{d_2}$. We conduct $L_2$ layers of convolution to the hypergraph and take each embedding of the last layer as the refined item embedding, i.e., $\mathbf{v}_i'=\mathbf{h}_i^{(L_2)}\in\mathbb{R}^{d_2}$.

\subsection{Frequency Based Gating}
\label{module:fbg}

In this section, we propose a gating mechanism based on user's interaction frequency to dynamically match the basket sequence to each item. This idea is built upon the prior knowledge that users tend to have a higher preference for those frequently interacted items.

Specifically, given the basket sequence $\mathcal{S}^u$, the probability of item $i$ containing in the next basket $b_{|\mathcal{S}^u|+1}^u$ is computed as \autoref{formula:output},
\begin{equation}
    \label{formula:output}
    \hat{y}_{\mathcal{S}^u,i}=\frac{1}{\sqrt{2d_2}}\left(\mathbf{W}_1^\top\left(\mathbf{v}_{\mathcal{S}^u}\oplus \mathbf{v}_i'\right)(1-\beta_i\alpha_i)+\gamma_i\alpha_i \right)
\end{equation}
where $\gamma^u\in\mathbb{R}^{|\mathcal{I}|}$ is the normalized vector whose $i$-th entry $\gamma^u_i$ represents the frequency of user $u$ interacting with item $i$. $\beta^u$ is the indicator vector of the positive entry in $\gamma^u$, i.e., $\beta^u=\mathbb{I}(\gamma^u>0)$, while gating vector $\alpha^u=\mathbf{W}_{2}\gamma^u+\mathbf{b}_2$. $\mathbf{W}_1\in\mathbb{R}^{2d_2\times1}$, $\mathbf{W}_2\in\mathbb{R}^{|\mathcal{I}|\times|\mathcal{I}|}$ and $\mathbf{b}_2\in\mathbb{R}^{|\mathcal{I}|}$ are learnable parameters. $\oplus$ represents the vector concatenate operation.

The recommendation loss is calculated as \autoref{formula:rec_loss},
\begin{align}
    \label{formula:rec_loss}
    \mathcal{L}_{Rec}=&-\frac{1}{|b_u^+|}\sum_{i\in b_u^+}\hat{y}_{\mathcal{S}^u,i}\log{y_{\mathcal{S}^u,i}}\nonumber\\
    &-\frac{1}{|\mathcal{I}/b_u^+|}\sum_{i\in\mathcal{I}/b_u^+}\left(1-\hat{y}_{\mathcal{S}^u,i}\right)\log\left(1-y_{\mathcal{S}^u,i}\right)
\end{align}
where $b_u^+$ denotes the ground-truth next basket following $\mathcal{S}^u$ and $y_{\mathcal{S}^u,i}$ denotes the label indicating whether $i\in b_u^+$.

\subsection{Recommendation and Training}
\label{module:loss}


For convenience, the size of the next basket $n$ is usually given as a hyperparameter by the algorithm. At the recommendation stage, we just recommend basket $\hat{b}_u$ to the user following \autoref{formula:recommend},
\begin{equation}
    \label{formula:recommend}
    \hat{b}_u=\underset{i\in\mathcal{I}}{\text{topn}}(\hat{y}_{\mathcal{S}^u,i})
\end{equation}
where $\text{topn}$ function returns a set of indices corresponding to the largest $n$ elements.

We optimize our HEKP4NBR model by a multi-task learning strategy as \autoref{formula:loss},
\begin{equation}
    \label{formula:loss}
    \mathcal{L}=\mathcal{L}_{PLM}+\mathcal{L}_{Rec}+\mathcal{L}_{B-I}+\mathcal{L}_{I-I}
\end{equation}

For clarity, we demonstrate the training procedure of HEKP4NBR in Algorithm~\ref{algo:training}.

\begin{algorithm}[!t]
	\caption{Training Procedure of HEKP4NBR}
	\label{algo:training}
	{\bfseries Input:} List of basket sequence $\mathcal{S}$, origin KG $\mathcal{KG}$, PLM pre-trained parameters $\Theta$, epoch number $T$.
	\begin{algorithmic}[1]
        \STATE Construct basket-item bipartite graph $\mathcal{G}_{B-I}$ according to $\mathcal{S}$.
		\STATE Randomly initialize the overhead parameters $\Phi$ and load $\Theta$ to PLM.
		\FOR{$i = 1,\dots, T$}
    		\FOR{$\forall \mathcal{S}^u\in \mathcal{S}$}
                \STATE Construct MUP by $\mathcal{S}^u$ via (\ref{formula:mup})
                \STATE Augment $KG$ and obtain KTP via (\ref{formula:aug_kg})-(\ref{formula:ktp})
                \STATE Obtain sequence embedding $\mathbf{v}_{\mathcal{S}^u}$ by PLM via (\ref{formula:seq_emb})
                \STATE Construct $\mathcal{HG}$ by $\mathcal{G}_{B-I}$ via (\ref{formula:moe1})-(\ref{formula:adj_matrix})
                \STATE Obtain refined item embeddings $\mathbf{v}_i'$ by hypergrpah convolutional module via (\ref{formula:hyper_conv})
                \STATE Predict probability by frequency based gating via (\ref{formula:output})
                \STATE Update model parameters $\Theta'$ and $\Phi'$ supervised by (\ref{formula:loss})
    		\ENDFOR
		\ENDFOR
	\end{algorithmic}
	{\bfseries Output:} PLM fine-tuned parameters $\Theta'$, overhead parameters $\Phi'$.
\end{algorithm}



\section{Experiments}

We conduct extensive experiments on two real world datasets to evaluate the effectiveness of the HEKP4NBR model. These experiments are designed to answering the following research questions:
\begin{itemize}
    \item \textbf{RQ1:} How does the proposed HEKP4NBR model perform compared to the state-of-the-art baselines in next-basket recommendation task?
    \item \textbf{RQ2:} What are the impacts of different key components and hyperparameters in HEKP4NBR?
    \item \textbf{RQ3:} What kind of PLM backbone is more suitable for HEKP4NBR?
    \item \textbf{RQ4:} How is HEKP4NBR's generalization to unseen templates?
\end{itemize}

\subsection{Experiments Settings}
\subsubsection{Datasets}

\begin{table}[]
\begin{tabular}{c|c|c|c}
\hline
                          & Statistics                     & Poultry  & Pharmacy \\ \hline
\multirow{5}{*}{Interactions}    & \# Users              & 11723 & 12946   \\
                                 & \# Items              & 238   & 6262    \\
                                 & \# Baskets            & 45746 & 63072   \\
                                 & Avg. basket size     & 4.133 & 3.366   \\
                                 & Avg. sequence length & 7.504 & 7.305   \\ \hline
\multirow{3}{*}{KG} & \# Relations          & 3     & 2       \\
                                 & \# Entities           & 253   & 6324    \\
                                 & \# Triplets            & 1428  & 25048   \\ \hline
\end{tabular}
\caption{Statistics of the datasets. Avg. basket size represents the average number of items in each basket. Avg. seq length represents the average length of basket sequence for each user.}
\label{tab:dataset}
\end{table}

We constructed two datasets based on the real sales data from two companies facing different types of customers. The first dataset, namely Poultry, consists of users' interactions with a large agribusiness in China. This company mainly provides B2B services, suggesting that users are highly adhesive wholesalers. The item names and attributes of Poultry dataset are all in Chinese which is not easy to translate into English. The second dataset, namely Pharmacy, is built upon the sales data of a medicine store in Shenzhen, China, which mainly focuses on B2C business. The names of the medicines are either professional medical terms or traditional Chinese herbal medicine names. 

We generate a basket sequence for each user by grouping every items interact at the same timestamp into a basket and then sorting the baskets in chronological order according to their timestamps. Following the well-known NBR method \cite{le2019correlation}, we filter out the baskets that contain fewer than 2 items and randomly select 5 items for those containing more than 5 items.Similarly, we filter out basket sequences whose length is less than 4 and keep the last 10 baskets for those longer than 10. Here, we set the maximum length of the basket sequence to 10, considering the limitation of the PLM input sequence's length. In Section~\ref{exp:hyperparam}, we will demonstrate the impact of maximum basket sequence lengths on overall performance. For each datasets, we divide all basket sequences into training set, validating set and testing set in the ratio of 8:1:1.

Since the datasets are based on non-public data, we manually build a knowledge graph for each dataset. For Poultry dataset, we connect each kind of poultry to its level, gender and category using relation \textit{level\_is}, \textit{gender\_is} and \textit{category\_is} respectively. For Pharmacy dataset, we connect each kind of medicine to its therapeutic function and category using relation \textit{function\_is} and \textit{category\_is} respectively. Some detailed statistics of two datasets after preprocessing are shown in \autoref{tab:dataset}.



\subsubsection{Evaluation Metrics}
Following \cite{zhai2023knowledge,M2}, we adopt three widely used metrics to evaluate the performance of our HEKP4NBR model, namely F1-score@$k$ (F1@$k$), Hit Rate@$k$ (HR@$k$) and Normalized Discounted Cumulative Gain@$k$ (NDCG@$k$). We report the metrics corresponding to $k=5$ and $k=10$ in the following section. Given $k$, we predict the next basket containing $k$ items for each user.

\subsubsection{Baselines}
To verify the effectiveness of our HEKP4NBR model, we choose the following state-of-the-art NBR methods as baselines in our experiments:
\begin{itemize}
    \item \textit{GRU4Rec} \cite{gru4rec} introduces Gating Recurrent Unit (GRU) to model the sequential dependency hidden in the users' interaction sequences. GRU4RecKG is a variant of GRU4Rec that incorporate KG to obtain better representation of items.
    \item \textit{STAMP} \cite{stamp} captures users' general interests by long-term attention and propose to model the current intention by a short-term memory mechanism.
    \item \textit{SASRec} \cite{sasrec} apply self-attentive encoder to the interaction sequences to generate the potential next item in a auto-regressive manner.
    \item \textit{NextItNet} \cite{nextitnet} encode the interaction sequence by a masked convolutional residual network composed of stacked 1 dimension dilated convolutional layers
    \item \textit{LightSANs} \cite{lightsans} assume that users' interests only concentrate in a constant number of aspects and generate the context-aware representation by the low-rank self-attention.
    \item \textit{$M^2$} \cite{M2} is a next-basket recommendation method which takes the users’ general preferences, items’ global popularities and transition patterns among items into consideration.
    \item \textit{GCSAN} \cite{gcsan} is a graph-based recommendation method that model the users' global interest by self-attention mechanism.
    \item \textit{SINE} \cite{sine} attempts to infer a set of interests for each user adaptively to model the current intention of users.
    \item \textit{BERT4Rec} \cite{bert4rec} is the first attempt to leverage Transformer based BERT encoder to encode the interaction sequence.
    \item \textit{P5} \cite{p5} is a PLM-based recommendation method that unified multiple recommendation tasks into a single framework by designing the personalized prompt sets.
\end{itemize}

Our baselines can be divided into three classes: (1) the models that focusing on modelling sequential patterns, i.e., GRU4Rec, GRU4RecKG, STAMP, SASRec, NextItNet, LightSANs, $M^2$, BERT4Rec and P5; (2) the models that based on graph, i.e., GCSAN and SINE.

\subsubsection{Implementation Details}
In our experiment, we utilize a pre-trained T5-small model \cite{t5} as the PLM backbone of HEKP4NBR. T5 is a text-to-text model based on encoder-decoder architecture whose encoder and decoder are both composed of 6 Transformer layers respectively with a hidden dimension $d_1=512$. Due to the length limitation of PLM, we set the maximum length of input tokens to 512 and only explore the entities within $n=3$ hops from the root entity $s$ when constructing KTP. We only set the number of GCN layer and hypergraph convolutional layer, i.e., $L_1$ and $L_2$, to 2 considering the oversmoothing problem occuring with too-deep GNNs. We set the number of experts in the MoE model to 8. For the embeddings dimensions, we set $d_2=128$, $d_3=64$. We randomly initialized the parameters in the overhead part of HEKP4NBR, i.e., the multi-item relation encoder and the frequency based gating module. We use 1 NVIDIA A800 GPU to train our model with a batch size of 64 and 100 epoch. We fully fine-tune the T5 backbone with a peak learning rate of 1e-5 and optimize the overhead with a peak learning rate of 1e-4 using AdamW optimizer.

\begin{table*}
\begin{tabular}{cc@{\hspace{0.65em}}c@{\hspace{0.65em}}c@{\hspace{0.65em}}c@{\hspace{0.65em}}c@{\hspace{0.65em}}c|c@{\hspace{0.65em}}c@{\hspace{0.65em}}c@{\hspace{0.65em}}c@{\hspace{0.65em}}c@{\hspace{0.65em}}c}
\hline
\multirow{2}{*}{Methods} & \multicolumn{6}{c|}{Poultry}                                                                           & \multicolumn{6}{c}{Pharmacy}                                                                         \\ \cline{2-13} 
                         & F1@5           & HR@5           & NDCG@5         & F1@10          & HR@10          & NDCG@10        & F1@5           & HR@5           & NDCG@5         & F1@10          & HR@10          & NDCG@10        \\ \hline
GRU4Rec                  & 0.451          & 0.579          & 0.574          & 0.374          & 0.769          & 0.661          & 0.172          & 0.231          & 0.327          & 0.116          & 0.258          & 0.338          \\
GRU4RecKG                  & 0.460          & 0.581          & 0.575          & 0.374          & 0.770          & 0.657          & 0.175          & 0.233          & 0.328          & 0.116          & 0.256          & 0.338          \\
STAMP                    & 0.418          & 0.537          & 0.518          & 0.357          & 0.736          & 0.609          & 0.171          & 0.229          & 0.326          & 0.114          & 0.254          & 0.337          \\
SASRec                   & 0.443          & 0.568          & 0.561          & 0.372          & 0.767          & 0.6548          & 0.173          & 0.233          & 0.327          & 0.117          & 0.260          & 0.337          \\
NextItNet                & 0.390          & 0.501          & 0.472          & 0.346          & 0.712          & 0.564          & 0.174          & 0.233          & 0.327          & 0.116          & 0.259          & 0.338          \\
LightSANs                & 0.440          & 0.561          & 0.547          & 0.366          & 0.746          & 0.6532          & 0.172          & 0.232          & 0.326          & 0.116          & 0.258          & 0.338          \\
BERT4Rec                 & 0.445          & 0.571          & 0.526          & \underline{0.391}          & {0.802}          & 0.631          & 0.179          & 0.244          & 0.327          & 0.118          & 0.268          & 0.336          \\
$M^2$                    & {0.486}          & {0.6520}          & \underline{0.594}          & 0.386          & 0.791          & {0.670}          & \underline{0.197}          & \underline{0.266}          & \underline{0.342}          & \underline{0.142}          & 0.315         & 0.362          \\
P5                       & \underline{0.510}               & \underline{0.635}               & 0.516               & 0.391               &  \underline{0.818}              & \underline{0.691}               & 0.184         & 0.246         & 0.341         & 0.142        & \underline{0.354}         &  \underline{0.376}    \\ \hline
GCSAN                    & 0.434          & 0.557          & 0.544          & 0.366          & 0.753          & 0.634          & 0.172          & 0.232          & 0.327          & 0.117          & 0.259          & 0.336          \\
SINE                     & 0.450          & 0.579          & 0.571          & 0.371          & 0.762          & 0.656          & 0.172          & 0.231          & 0.328          & 0.116          & 0.259          & 0.339          \\ \hline
HEKP4NBR                     & \textbf{0.627} & \textbf{0.793} & \textbf{0.807} & \textbf{0.439} & \textbf{0.896} & \textbf{0.783} & \textbf{0.273} & \textbf{0.368} & \textbf{0.447} & \textbf{0.170} & \textbf{0.377} & \textbf{0.428} \\ 
Improv.                  & 22.94\%        & 24.88\%        & 35.85\%        & 12.27\%        & 9.54\%        & 13.31\%        & 38.57\%        & 38.34\%        & 30.70\%        & 19.71\%        & 6.49\%        & 13.82\%       \\ \hline
\end{tabular}
\caption{Overall performance. Bold scores represent the highest results of all methods. Underlined scores stand for the second highest results of all methods.}
\label{tab:comparison}
\end{table*}

\subsection{Performance Comparison (RQ1)}
\label{exp:comparison}

The performance comparison results of all baselines and the proposed HEKP4NBR model are shown in \autoref{tab:comparison}. According to the experimental results, we may conclude the following observations.

The proposed HEKP4NBR model demonstrates considerable performance improvements across all evaluation metrics and datasets. On one hand, improvements in NDCG metric suggest that our method can not only accurately predict items in the next basket, but also demonstrates a strong comprehensions on the significance and relevance between the next basket and each potential item. This is likely attributed to the prior knowledge encoded in the PLM corpus and KG, which empowers our model to alleviate the OOV item issue and obtain better representations of items. On the other hand, our method also achieves competitive improvements in the HR metric, indicating its capability to cover a large portion of ground truth items in the next basket. We believe that this improvement stems from the modelling of correlation among multiple items by the multi-item relation encoder. It's worth noting that P5 achieves almost the highest performance among all baselines, solely utilizing PLM to encode the prompt text corresponding to the interaction sequence without using any additional side information. Nevertheless, P5 requires 300 epochs and about twice as much training time as HEKP4NBR to converge. In light of this, we argue that using MUP to transform recommendation task into a pre-trained mask prediction task significantly contributes to the convergence speed of model. 

Some baselines (GRU4Rec, SASRec, BERT4Rec) attempt to precisely leverage sequential encoders, such as GRU, self-attention mechanism and Transformers, to accurately model the transition patterns from basket to basket in the interaction sequence for each user. While these methods have produced competitive results, relying solely on item IDs leads to the model treating each item independently. Consequently, the model can only capture shallow ID transition patterns and lack the ability to discern deeper semantic-level correlations. Some baselines like GRU4RecKG make efforts to integrate side information such as KG at an early stage. These methods indeed benefit to boost performance but still cannot achieve the best results. This is mainly because obtaining high quality item representations in large scale KG is a challenging task due to intrinsic noise in the KG. Furthermore, the pre-trained item representations may not be suitable for the recommendation task. Unlike the trivial way in GRU4RecKG, HEKP4NBR leverages KG at a finer granularity by transforming the most relevant knowledge into KTP. In this way, we are able to model the deep semantic correlations while alleviating the effects of noise in KG by exploiting the excellent comprehension capabilities of PLM. We also notice that $M^2$ can achieve better results than GRU4RecKG even without any prior knowledge to items. This is because $M^2$ is the only baseline that explicitly model user preference. In Section~\ref{exp:ablation}, we will show that modelling user preference by simply counting purchase frequency will significantly boost performance.

Other baseline (GCSAN, SINE) introduce graph structure to facilitate sequence modelling and thus achieves relatively competitive performance. Specifically, SINE mine different aspects of user interest from the user's session graph and performs better compared with GCSAN in Poultry dataset. This indicates that mining multiple patterns from the interaction sequence will benefit the model, which shares the same idea with MoE component in HEKP4NBR.

\subsection{Ablation Study (RQ2)}
\label{exp:ablation}
\begin{table}[]
\begin{tabular}{c|clll|ccc}
\hline
Datasets                  & \multicolumn{4}{c|}{Settings}      & F1@5           & HR@5           & NDCG@5         \\ \hline
\multirow{5}{*}{Poultry}     & \multicolumn{4}{c|}{w/o GCN}       & 0.453          & 0.577          & 0.545          \\
                          & \multicolumn{4}{c|}{w/o Hyper-GCN} & 0.547          & 0.692          & 0.684          \\
                          & \multicolumn{4}{c|}{w/o FBG}       & 0.471          & 0.591          & 0.605          \\
                          & \multicolumn{4}{c|}{w/o KTP}       & 0.511          & 0.649          & 0.632          \\
                          & \multicolumn{4}{c|}{HEKP4NBR}          & \textbf{0.627} & \textbf{0.793} & \textbf{0.807} \\ \hline
\multirow{5}{*}{Pharmacy} & \multicolumn{4}{c|}{w/o GCN}       & 0.234          & 0.319          & 0.392          \\
                          & \multicolumn{4}{c|}{w/o Hyper-GCN} & 0.212          & 0.291          & 0.363          \\
                          & \multicolumn{4}{c|}{w/o FBG}       & 0.190          & 0.258          & 0.348          \\
                          & \multicolumn{4}{c|}{w/o KTP}       & 0.238          & 0.325          & 0.398          \\
                          & \multicolumn{4}{c|}{HEKP4NBR}          & \textbf{0.273} & \textbf{0.368} & \textbf{0.447} \\ \hline
\end{tabular}
\caption{Ablation study of HEKP4NBR and its four variants on Poultry (above) and Pharmacy (below) Dataset. Bold scores represent the highest results of all variants. }
\label{tab:ablation}
\end{table}

We validate the effectiveness of several components of HEKP4NBR by designing ablation study on two datasets. The setting of each variant is described below:
\begin{itemize}
    \item w/o GCN: We remove the GCN part to obtain initial item embeddings $\mathbf{v}_i$ and randomly initialize $\mathbf{v}_i$ instead.
    \item w/o Hyper-GCN: We remove the hypergraph convolutional module and MoE model and directly feed $\mathbf{v}_i$ into the FBG module.
    \item w/o FBG: We replace the dynamic gating module with a simple inner product operation between $\mathbf{v}_{\mathcal{S}^u}$ and $\mathbf{v}_i'$, that is,
    \begin{equation}
        \hat{y}_{\mathcal{S}^u,i}=v_{\mathcal{S}^u}\cdot v_i'
    \end{equation}
    \item w/o KTP: We remove the KTP and only feed MUP into PLM. 
\end{itemize}

The results of ablation study are shown in \autoref{tab:ablation}. We can see that the performance degenerate in different degrees of all four variants, which verifies the effectiveness of our components. Specifically, it's evident that on the Poultry dataset, which comprises relatively fewer items and denser interactions, the lack of GCN and FBG will significantly deteriorate performance, resulting in a degradation of more than 30\%. This indicates that utilizing collaborative signal between baskets and items in the basket-item bipartite graph $\mathcal{G}_{B-I}$ can enhance the quality of item embeddings. 
Similarly, a smaller item set implies denser purchase frequency vectors $\gamma$ of users, hence implying that users have more distinct preferences and re-purchase patterns. In this case, the proposed FBG is more powerful in mining these patterns. 

On the contrary, on the larger and sparser Pharmacy dataset, the hypergraph convolutional module plays an important role, which will cause over 25\% degradation if removed. This is because there are more combinations of items and preferences vary among users in Pharmacy dataset while the hypergraph convolutional module is suitable for modelling semantic correlations of different item combinations. 

We also find that leveraging KG as side information can improve performance by approximately 20\% and 10\% for the Poultry and Pharmacy datasets, respectively. We can conclude that the proposed KTP can enable our model to mine the deep semantic information of items and thus boost performance especially for long tail items and cold start items.

\subsection{Hyperparameter Analysis (RQ2)}
\label{exp:hyperparam}

\begin{figure}
	\centering
		\includegraphics[width=0.8\linewidth]{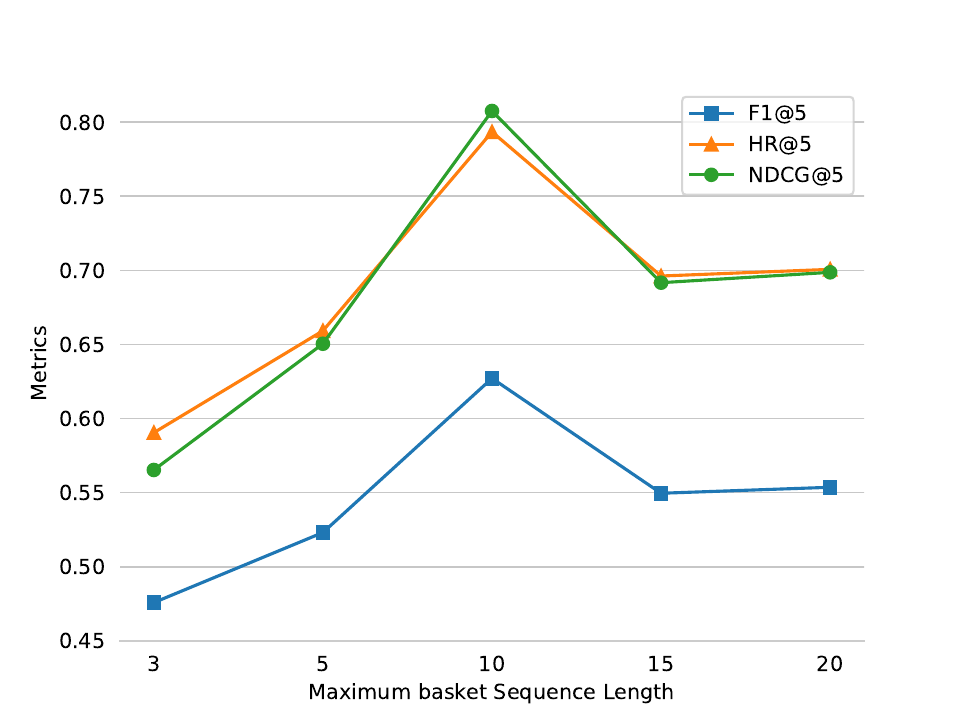}
		\caption{Performance with different maximum basket sequence length while fixing PLM input token sequence length on Poultry dataset.}
	\label{fig:max_seq_len}
\end{figure}

In Section~\ref{exp:comparison} and Section~\ref{exp:ablation}, we have shown that transforming KG into KTP and feeding KTP together with MUP into PLM can benefit the recommendation performance. However, since the input token sequence length of PLM cannot be infinite, how to balance between the length of MUP and KTP becomes a trade-off problem and we are trying to investigate the reasonable proportion between the two lengths in this section. 

We first fix the maximum input token sequence length to 512 tokens and then change the maximum basket sequence length to 3, 5, 10, 15 and 20 respectively, noting that a longer basket sequence means that a longer MUP and thus a shorter KTP. We examine the overall performance corresponding to different maximum basket sequence length and report the results on Poultry dataset in \autoref{fig:max_seq_len}.

We can see that all of the three metrics peak when the maximum basket sequence length is set to 10, which is consistent with our setting in the previous study. As the maximum basket sequence length decreases, all three metrics decline sharply, even though more side information is fed into PLM. This implies that the importance of user's interaction history outweighs the prior knowledge of items. When the interaction history is incomplete, the model exhibits significant bias in modelling user preferences. However, a longer basket sequences do not necessarily guarantee better performance. This is because user preferences often change over time. Recent interactions tend to represent the user's latest preferences more accurately while interactions from too long ago will introduce noise into the model. Additionally, MUP that is too long can result in insufficient knowledge input to PLM, leading to performance degradation.

\subsection{Impact of PLM (RQ3)}

\begin{table}[]
\begin{tabular}{c@{\hspace{0.5em}}c@{\hspace{0.5em}}c@{\hspace{0.6em}}c@{\hspace{0.6em}}c}
\hline
PLM     & Architecture    & Layers & Dimension & Parameters \\ \hline
T5 \cite{t5}      & encoder-decoder & 6+6                & 512              & 60M       \\
BART \cite{bart}    & encoder-decoder & 6+6                 & 768              & 140M      \\
BERT \cite{bert}    & encoder-only    & 12                & 768              & 110M      \\
DeBerta \cite{deberta} & encoder-only    & 12                & 768              & 100M      \\ \hline
\end{tabular}
\caption{Basic information of four PLMs. Layers $a+b$ in encoder-decoder PLMs means $a$ Transformer layers in encoder and $b$ layers in decoder. }
\label{tab:plm}
\end{table}

\begin{figure}
	\centering
		\includegraphics[width=0.75\linewidth]{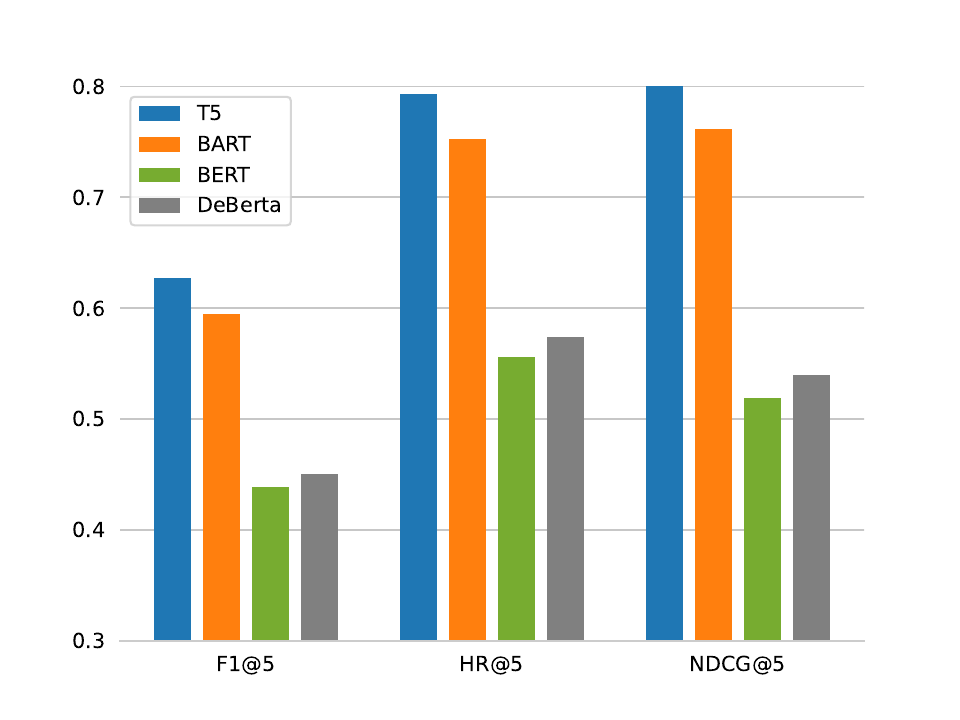}
		\caption{Performance with different PLM backbones on Poultry dataset.}
	\label{fig:plm}
\end{figure}

We wonder what kind of PLM is more suitable for HEKP4NBR, thus we examine the performance based on four PLM backbones which can be divided into two architectures: encoder-decoder (T5, BART) and encoder-only (BERT, DeBerta). Some basic information of four PLMs is listed in \autoref{tab:plm}. The results on Poultry dataset are shown in \autoref{fig:plm}. We can conclude the following observations from the results.

First of all, the encoder-decoder architecture PLMs significantly surpass those of the encoder-only architecture. This is because encoder-decoder PLMs perform better at chronological modelling and therefore are more suitable for sequence-to-sequence generation tasks such as mask prediction, translation and question answering. On the contrary, encoder-only PLMs are designed to extract global contextual features of the input sequence and thus perform better on global comprehension tasks like sentence classification.

What's more, large number of parameters is not sufficient for good performance, while the gap between pre-train task and downstream task has a significant impact. Although BART has more parameters than T5, it fails to surpass in performance. This is because BART is pre-trained by reconstructing documents corrupted by text infilling and sentence shuffling \cite{bart}, while T5 is pre-trained on BERT-style mask prediction task \cite{t5} which is perfectly suited to the recommendation task. Therefore, we are able to bridge the gap between the T5's pre-train task and downstream task by MUP.

\subsection{Generalization on unseen templates (RQ4)}

\begin{figure}
	\centering
		\includegraphics[width=0.75\linewidth]{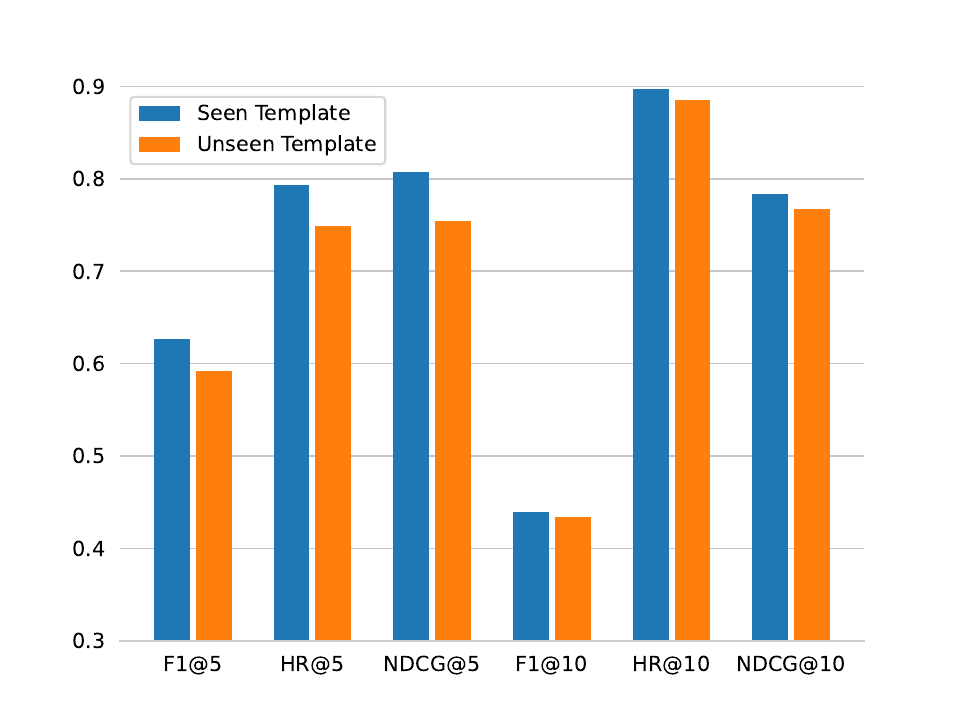}
		\caption{Generalization performance of HEKP4NBR when testing on seen and unseen template on Poultry dataset.}
	\label{fig:template}
\end{figure}

In the previous study, we use the same MUP and KTP template for all basket sequences in the training, validating and testing sets. In this section, we want to determine whether HEKP4NBR is sensitive to the designed template. We manually design 3 different templates for the Poultry dataset and train, validate and test HEKP4NBR using one template each. The performance of HEKP4NBR when testing on seen and unseen template is reported in \autoref{fig:template}. We can see that the performance only slightly declined on unseen template by 5\% at most, indicating that our method can generalize across different templates.


\section{Conclusion}

In this paper, we propose HEKP4NBR, which is a hypergraph enhanced PLM-based NBR method. Firstly, we transform KG into KTP to help PLM encode the OOV item IDs in the user's basket sequence. In this way, we are able to transform NBR task into mask prediction pre-train task following the prompt learning paradigm. Secondly, we proposed a novel hypergraph convolutional module. Specifically, we build a hypergraph based on item similarities measured by an MoE model from multiple aspects and employ convolution on hypergraph to model correlations among multiple items. Finally, extensive experiments demonstrate the effectiveness of HEKP4NBR. In future work, we will continue to explore how to use large scale PLMs to boost performance in a parameter efficient way. 

\section{Acknowledgments}

\bibliographystyle{ACM-Reference-Format}
\bibliography{sample-base}










\end{document}